\documentclass[amsfonts,amsmath,,usenames]{article}
\usepackage{jheppub}

\usepackage{mathpazo}   % palatino w/mathpazo fonts for math
\usepackage[tight,nice]{units}

\newcommand {\be} {\begin{equation}} 
\newcommand {\ba}{\begin{eqnarray}} 
\newcommand {\ee} {\end{equation}} 
\newcommand{\ea} {\end{eqnarray}}

\begin{document}

\title{Asymptotic Freedom in Holographic QCD}

\author{Zainul Abidin,} 
\author{Herry J. Kwee}
 \author{and Jong Anly Tan}
\affiliation{Theoretical Physics Group, Physics Department, STKIP Surya,\\
Gedung SURE Jl. Scientia Boulevard Blok U/7 Gading Serpong, Tangerang 15810, Banten, Indonesia}

\date{\today}

\abstract{
We calculate the running of the quark mass and the quark condensate using a dynamical soft-wall model by Csaki and Reece. We show that a correct running can be obtained with an appropriate sign for the dilaton field. In the soft-wall model with quadratic dilaton profile, a wrong sign for the dilaton field can give rise to  a massless vector meson, a signature of spontaneously broken symmetry in the vector sector, which is not observed in nature. With a right parameter range, we obtain no such massless vector meson. We also find that, contrary to the soft-wall model with quadratic dilaton profile, the model allows a non-vanishing quark condensate in the chiral limit. We obtain the vector meson mass spectrum similar to that of the hard-wall model. By varying a parameter in the model we can fit the running of the quark mass and of the quark condensate.
%allows the explicit symmetry breaking to be independence from the spontaneous symmetry breaking.
%is free from the problem which plague the soft-wall model with quadratic dilaton profile, which is the vanishing of the quark condensate in the chiral limit.
}
\emailAdd{zainul.abidin@stkipsurya.ac.id}
 \emailAdd{herry.kwee@stkipsurya.ac.id}
  \emailAdd{jongtan@stkipsurya.ac.id}
\keywords{AdS/CFT, holography, QCD, asymptotic freedom}

\maketitle

%%%%%%%%%%%%%%%%%%%%%%%%%%%%%%%%%%%%%%%%%%%%

\section{Introduction}			\label{sec:one}

%%%%%%%%%%%%%%%%%%%%%%%%%%%%%%%%%%%%%%%%%%%%
Understanding Quantum Chromodynamics (QCD) at the low-energy region has been a challenge for  
physicists in the past half century.  While the high-energy region of QCD is asymptotically free~\cite{Gross:1973id} and can be studied and understood well using perturbative method, in the low-energy region, QCD interactions actually become strong and non-linear and cannot be studied perturbatively. 
%This behavior may also explain the quark-confinement phenomena (nobody has ever succeeded to isolate a free quark).  
Another challenge is to understand the structure of nucleons as the non-linearity 
of QCD cause the interactions between quarks in the low-energy region to 
be very complicated.

For the lack of formal derivation, most of the efforts to understand QCD at
low-energy region are concentrated in building models that closely resemble
QCD at this energy scale, in particular in modeling hadrons. Many models have been proposed in the past, 
such as Bag Model~\cite{Johnson:1985nh}, Constituent Quark Model
~\cite{Capstick:1986bm}, Chiral Soliton Model~\cite{Skyrme:1962vh}, Chiral 
Lagrangian~\cite{Coleman:1969sm, Gasser:1983yg}, etc.  While most of these models work quite well in the energy 
region they are prescribed and explain certain properties and behavior of 
hadrons, they are not derived from first principle. They also 
generally cannot extend to the high-energy region where QCD works very well.

 In 1997 Maldacena conjectured that a certain type of string theory, in particular the type IIB string theory  in AdS$_5\times$S$_5$
corresponds to the conformally invariant large N$_c$ SU(N$_c$) ${\cal N}=4$ 
Supersymmetric theory on the 4-dimensional space on the surface of the 
5-dimensional AdS space~\cite{Maldacena:1997re}. This is how the moniker AdS/CFT got its name. With this 
conjecture, the relationship becomes more tangible and a dictionary can be composed 
to relate parameters and symmetries in one theory to another~\cite{Gubser:1998bc, Witten:1998qj}. Better yet, the two 
regions of the  theory cross where strong coupling, {\it i.e.} non-linear region 
in one theory corresponds to weak coupling, {\it i.e.} perturbative region in another 
theory. To calculate a theory in the strong region all we have to do is to 
calculate the corresponding theory in the perturbative region.
%AdS/CFT is a conjectured duality between a weakly coupled gauge theory in 4D, and a strongly coupled gravity theory in 5D~\cite{Maldacena:1997re, Gubser:1998bc, Witten:1998qj}. The duality allow us to access the strongly coupled regime of a gauge theory via its gravitational dual theory. Especially attractive application of the duality is in accessing the low energy limit of QCD through a perturbative expansion in the dual theory. 
There are two approaches  in constructing the dual of QCD, dubbed the top-down and the bottom-up approaches. In the former, one starts from a string theory with an appropriate brane configuration~\cite{Kruczenski:2003uq, Sakai:2004cn, Sakai:2005yt}. In the latter, one begins by introducing a 5D action with the few fields needed to reproduce the well-known properties of low-lying QCD states~\cite{Erlich:2005qh, DaRold:2005zs, Hirn:2005nr}. In this paper we will follow the bottom-up approach.

In its simplest form, the bottom-up models use a 5D AdS metric with cutoff to break conformal symmetry and simulate confinement in QCD. The 5D fields present in the action are bifundamental scalar field $X(x,z)$ and gauge fields, $L(x,z)$ and $R(x,z)$, satisfying the SU$(2)_{\rm L} \times$ SU$(2)_{\rm R}$ symmetry.   These 5D fields corresponds to $\bar q_L q_R(x)$, $J_L(x)$ and  $J_R(x)$ operators in the 4D gauge theory respectively. The model produces QCD features quite well~\cite{Grigoryan:2007vg, Kwee:2007dd, Grigoryan:2007wn, Kwee:2007nq, Erlich:2008en, Abidin:2008ku, Abidin:2008hn, Abidin:2009aj, Erlich:2009me}. However, the model has some drawbacks. There are ambiguities in the infrared boundary conditions, and the resulting meson spectrum does not satisfy the observed Regge trajectory where the mass squared grows as $n$. Instead, the spectrum of the mass is similar to that of the quantum system of a particle in a box, {\it i.e.} $m_n^2$ grows as $n^2$.

Another way to simulate confinement is by introducing a background dilaton field to the 5D action~\cite{Karch:2006pv}. In order to obtain a meson spectrum that satisfies linear Regge trajectory, the profile of the dilaton field is chosen to be quadratic while the metric of the 5D space remains AdS. However, while the soft-wall model successfully reproduce the linear Regge trajectory, it fails to reproduce a QCD-like chiral symmetry breaking.  In particular, in the limit of vanishing quark mass parameter, the quark condensate parameter is vanishing as well.  A way to overcome this problem, among other things, was pointed out in \cite{Gherghetta:2009ac}. They introduced a quartic term to the bulk scalar potential,  included a tachyon field and modified the dilaton profile.

A dilaton field with opposite sign, relative to the one used in~\cite{Karch:2006pv}, has been considered in the literature \cite{deTeramond:2009xk, Zuo:2009dz, Nicotri:2010at} as it seems to have more appealing confinement properties. In~\cite{Karch:2006pv} and later reemphasized in~\cite{Karch:2010eg}, this reversed sign choice was discarded because it leads to an unphysical massless vector meson.  It was shown that the problem can not be easily remedied unless the asymptotic value of the dilaton field at the IR boundary is changed. With the correct dilaton sign, the wave function of the massless mode is non-normalizable, hence, decouple. 

%Various aspect of QCD has been investigated using the holographic principle. We will mention few of them here. Relationship between  parameters in AdS/QCD and parameters of QCD in light-front frame has been discussed in~\cite{Brodsky:2003px, Brodsky:2006uqa, Brodsky:2007hb}. QCD with finite baryon number density in AdS/QCD has been investigated in~\cite{Domokos:2007kt} and for finite isospin number density in~\cite{Albrecht:2010eg}. The axial anomaly is discussed in~\cite{Grigoryan:2008up}. The confinement/deconfinement phase transition has been considered in. 

Various aspect of QCD has been investigated using the holographic principle~\cite{Katz:2005ir, Andreev:2006ct, Brodsky:2003px, Brodsky:2006uqa, Brodsky:2007hb, Herzog:2006ra, BallonBayona:2007vp, Domokos:2007kt, Grigoryan:2008up, Albrecht:2010eg, Vega:2010ns}. In this paper, we will use a model introduced in~\cite{Csaki:2006ji} where back reaction by the dilaton field to the AdS metric is considered.  Even in the simplest case in which no potential term for the dilaton field presents, the back reaction from the dilaton field automatically provides an IR cutoff. With an addition of appropriate potential terms for the dilaton field, the model can reproduce asymptotic freedom of QCD quite well. 

The organization of this paper is as follows. In section 2, we will review the original AdS/QCD proposal for the hard-wall and the soft-wall model. In section~\ref{sec:incorp}, we will review the holographic model in~\cite{Csaki:2006ji}. We will discuss the vector part of the action in section~\ref{sec:four}, and show that the spectrum does not contain massless vector meson. In section~\ref{sec:three}, we consider the scalar part of the action and investigate the running of the quark mass and the quark condensate. Finally in section~\ref{sec:concl}, we provide the conclusions.

%%%%%%%%%%%%%%%%%%%%%%%%%%%%%%%%%%%%%%%%%%%%
%
\section{AdS/QCD}
%
%%%%%%%%%%%%%%%%%%%%%%%%%%%%%%%%%%%%%%%%%%%%
In AdS/CFT there is a correspondence between 5-dimensional fields $\phi(x,z)$, and 4-dimensional operators $\mathcal{O}(x)$ living on the boundary of the 5-dimensional Anti-de Sitter (AdS$_5$) space,  where $z$ is the fifth coordinate. The metric of AdS$_5$ can be written as
\be
ds^2=\frac{R^2}{z^2}\left(\eta_{\mu\nu} dx^\mu dx^\nu-dz^2\right)\,.
\ee
The fifth coordinate, $z$, corresponds to energy scale (or momentum transfer $Q^2$), {\it i.~e.}, $z\to 0$ corresponds to infinite energy ($Q^2 \to \infty$). 

The 4D sources in the 4D generating functional $Z_{\rm 4D}[\phi^0]$ is called $\phi^0(x)$ and the generating functional is defined as follow:
\be
Z_{\rm 4D}[\phi^0]=\left<\exp\left(i\int d^4x \phi^0(x)\mathcal{O}(x)\right)\right>\,.
\ee
The correspondence maybe written as
\be
Z_{\rm 4D}=e^{iS_{\rm 5D}[\phi_{cl}]}\,,
\ee
where on the right hand side $S_{\rm 5D}[\phi_{cl}]$ is the action evaluated for the classical solution $\phi_{cl}(x,z)$ to the field equations with boundary condition
\be
\lim_{z\to0} \phi_{cl}(x,z)=z^{\Delta} \phi^0(x)\,.
\ee
The constant $\Delta$ depends on the nature of the operator $\mathcal{O}$. Utilizing the correspondence, one obtain the $n$-point correlator
\be
\left<0|\mathcal{T} \mathcal{O}(x_1)\ldots \mathcal{O}(x_n)|0\right>=\frac{(-i)^n\delta^n e^{iS_{5D}}}{\delta\phi^0(x_1)\ldots\delta\phi^0(x_n)}\bigg|_{\phi^0\to 0}\,.
\ee

There are an infinite number of 5D fields corresponding to an infinite number of 4D operators in QCD. However, there are only several crucial operators involved in the dynamics of spontaneous chiral symmetry breaking. One of these crucial operators is the quark bilinear $\bar\psi_L\psi_R$, whose non-vanishing vacuum expectation value breaks the full symmetry  SU$(N_f)_L\times$SU$(N_f)_R$ down to SU$(N_f)_V$. The corresponding 5D field of the quark bilinear operator is a scalar field, $X$, which transform as a bifundamental. The conserved current operators of the SU$(N_f)_L\times$SU$(N_f)_R$ symmetry, $J^{a\mu}_L(x)$ and $J^{b\nu}_R(x)$, correspond to the 5D gauge field $L^a_\mu(x,z)$ and $R^b_\nu(x,z)$, respectively. 

\subsection{Hard-wall}
In the hard-wall model~\cite{Erlich:2005qh}, where the 5-dimensional space is AdS with cutoff, the meson part of the action can be written as
\be
{\cal S} = \int d^5x \sqrt{g} 
  {\rm Tr} \lbrace |DX| + 3|X|^2 - \frac{1}{4g_5^2} (F_L^2 + F_R^2) \rbrace, \label{mesonhardaction}
\ee
with $D^M X = \partial^M X - iA_L^M X + i XA_R^M$ and $F_{L,R}^{MN} = \partial^M A_{L,R}^N - \partial^N A_{L,R}^M - i[A_{L,R}^M,A_{L,R}^N]$. The field $X$ can be written as  a product of a background scalar field $X_0(z)$ and a pseudo-scalar field $\pi^a(x,z)$
\be
X(x,z)=X_0(z)\exp(i2\pi^at^a).
\ee
In the case of $N_f=2$ and assuming isospin symmetry, one can take $X_0$ as multiple of the identity, and one does not have to worry about the ordering. The solution to the background scalar field equation of motion can be written as
\be
X_0(z)=\frac{1}{2}\left(m_q \zeta z + \frac{1}{\zeta}\sigma_q z^3\right)\,, \label{mqsigEq}
\ee
where the constant $m_q$ can be identified as the quark mass sourcing the bilinear operator $\bar\psi_q\psi_q$ and the constant $\sigma_q$ as the vacuum expectation value of the quark bilinear operator. The constant $\zeta=\sqrt{ N_c}/2\pi$ was discussed in~\cite{DaRold:2005vr, Cherman:2008eh}.

\subsection{Soft-wall}
In the soft-wall model~\cite{Karch:2006pv}, confinement is modeled with a background dilaton field $\phi(z)=-\kappa z^2$. This dilaton profile is chosen so that the meson spectrum behaves like, $m_n^2\sim n+S$, for excitation number $n$ and spin $S$. The dilaton field modifies the Lagrangian (\ref{mesonhardaction}) in the form of an additional exponential factor $e^{\phi}$. The meson part of the action can be written as
\be
{\cal S} = \int d^5x\, e^{\phi}\,\sqrt{g} 
  {\rm Tr} \lbrace |DX| + 3|X|^2 - \frac{1}{4g_5^2} (F_L^2 + F_R^2) \rbrace, \label{mesonsoftaction}
\ee
The scalar field satisfies
\be
\partial_z^2X_0 -\left(2\kappa z +\frac{3}{z}\right)\partial_z X_0 +\frac{3}{z^2}X_0=0\,,
\ee
with the following solution
\be
X_0(z)=c_1 z U\left(\frac{1}{2},0,\kappa z^2\right)+c_2 z^3 M\left(\frac{3}{2},2,\kappa z^2\right)\,,
\ee
where $U$ dan $M$ are Kummer's functions. However, in order that the action evaluated on the solution is finite, the second term in the above solution must be dropped. Expanding the remaining term in the small $z$ limit, one obtains
\be
X_0(z)= \frac{2c_1}{\sqrt{\pi}}\,z + \frac{c_1\kappa}{\sqrt{\pi}}\left(1+\gamma_E +\ln(\kappa z^2/4)\right) z^3\,,
\ee
where $\gamma_E$ is the Euler-Mascheroni constant. Identifying $c_1=m_q \zeta\sqrt{\pi}/4$, the above solution leads to $\sigma\propto m_q$. In the limit $m_q\to 0$, we eliminate both the explicit and the spontaneous symmetry breaking, in contradiction with QCD.

In this paper, we use a model in which the dilaton field is dynamic. In the next section, following~\cite{Csaki:2006ji}, we will choose a profile for the dilaton field that produces asymptotic freedom of QCD and obtain the potential term associated with the dilaton profile. We find that the model allows for a non-vanishing quark condensate in the limit of $m_q\to 0$.

%%%%%%%%%%%%%%%%%%%%%%%%%%%%%%%%%%%%%%%%%%%%
 
  \section{Incorporating Asymptotic Freedom}
 \label{sec:incorp}
 
 %%%%%%%%%%%%%%%%%%%%%%%%%%%%%%%%%%%%%%%%%%%%
The goal of AdS/QCD is to reproduce a model that resembles QCD as close as 
possible, including QCD asymptotic freedom behaviour.  Following Csaki 
{\it et. al.} formulation~\cite{Csaki:2006ji}, we define our action
\begin{equation}
 {\cal S}_{g\phi} = \frac{1}{2\kappa^2} \int d^5x \sqrt{g} \left(-{\cal R} -
 V(\phi) + \frac{1}{2} g^{MN} \partial_M \phi \partial_N \phi \right),
\end{equation}
with $\kappa^2$ the 5-dimensional Newton constant and $V(\phi)$ chosen to 
reproduce asymptotic freedom. Note that $\phi$ is dimensionless here.

We assume QCD gauge coupling is given by $e^{b\phi(z)}$ just like in string 
theory. For the coupling to run logarithmically we need a solution of the
form:
\begin{equation}
 e^{b\phi(z)} = \frac{1}{\log \frac{z_0}{z}}, \label{bphi}
\end{equation}
where as usual we have identified the AdS coordinate $z$ with the energy 
scale and defined $z_0=\Lambda^{-1}_{QCD}$.  We do not fix the value of 
$b$ {\it a priori}. It is easier to work in coordinate $y$ given by 
$\exp{(y/R)}\equiv z/R$ in which our solution will take the form: $e^{b\phi(y)} 
= \frac{R}{y_0-y}$. Inverting this we obtain the function for the dilaton 
\begin{equation}
 \phi(y) = \frac{1}{b} \log{\left(\frac{R}{y_0-y}\right)},
\end{equation}
and its derivative
\begin{equation} 
 \phi^\prime(y) = \frac{1}{b(y_0-y)} = \frac{1}{bR} e^{b\phi}.
 \label{eq:dilaton}
\end{equation} 

First we write our metric in the $y$ coordinate:
\begin{equation}
 ds^2= e^{-2A(y)} (\eta_{\mu\nu} dx^\mu dx^\nu - dy^2).
\end{equation}
Subtituting the metric into our action we obtain:
\begin{equation}
 {\cal S}_{g\phi} = \frac{1}{2\kappa^2} \int d^5x \sqrt{g} \left( -20 A^{\prime 2} 
  +8 A^{\prime\prime} - V(\phi) - \frac{1}{2} (\phi^\prime)^2 \right).
\end{equation}
Taking the variance of this action with respect to $A$ and $\phi$ and setting
it to zero:
\begin{eqnarray}
 \frac{\delta S_{g\phi}}{\delta A} &=& 0 \to 48 A^{\prime 2} - 24 A^{\prime \prime} 
 + 4V(\phi) + 2 (\phi^\prime)^2 = 0, \nonumber \\
 \frac{\delta S_{g\phi}}{\delta \phi} &=& 0 \to \phi^{\prime\prime} 
 - 4 \phi^\prime A^\prime = \frac{\partial V}{\partial\phi}.
 \label{eq:action_var}
\end{eqnarray}

Next we use superpotential method by defining a function $W(\phi)$ such that:
\begin{eqnarray}
 A^{\prime}(y) &=& W(\phi(y)), \nonumber \\
 \phi^\prime(y) &=& 6\frac{\partial W}{\partial\phi}. 
\end{eqnarray}
Solving for $W(\phi)$ with $\phi^\prime$ as defined in Eq.~\ref{eq:dilaton}, we 
obtain: 
\begin{equation}
 W(\phi) = \frac{1}{6b^2R} e^{b\phi} + W_0.
\end{equation}

It is very easy to show that the potential is
\begin{equation}
 V(\phi) = 18 \left(\frac{\partial W}{\partial\phi}\right)^2 - 12W^2.
\end{equation}
Now we can solve for the warp factor:
\begin{equation}
 A(y) = A_0 + W_0 y + \frac{1}{6b^2} \log{\frac{R}{y_0-y}},
\end{equation}
or in $z$ coordinate:
\begin{equation}
 A(z) = A_0 + W_0 R\log{\frac{z}{R}} -\frac{1}{6b^2} \log{\log{\frac{z_0}{z}}}.
\end{equation}
For $e^{-2A(z)} = e^{-2A_0} \left(\frac{R}{z}\right)^{2W_0R} 
\left(\log{\frac{z_0}{z}}\right)^{\frac{1}{3b^2}}$ to resemble AdS metric 
with $\log z_0/z$ factor, we need $A_0=0$ and $W_0=1/R$. This will give us
\ba
 W(\phi)& =& \frac{1}{R} \left(\frac{1}{6b^2}e^{b\phi}+ 
  1 \right), \\
   V(\phi)&=&  -\frac{1}{3b^2R^2} \left(\left(\frac{1}{b^2} - \frac{3}{2}\right) 
  e^{2b\phi} + 12e^{b\phi} + 36b^2 \right)
\ea
and  
\begin{equation}
 A = \frac{y}{R} + \frac{1}{6b^2}\log{\frac{R}{y_0-y}} = 
  \log{\frac{z}{R}} - \frac{1}{6b^2}\log{\log{\frac{z_0}{z}}}.
\end{equation}
For the case $b=\pm\sqrt{\frac{2}{3}}$, we got:
\begin{equation}
 V(\phi) = -\frac{6}{R^2}e^{\pm\sqrt{\frac{2}{3}}\phi} - \frac{12}{R^2},
\end{equation}
and  
\begin{equation}
 \phi = \mp\sqrt{\frac{3}{2}} \log{\frac{y_0-y}{R}} = 
  \mp\sqrt{\frac{3}{2}} \log{\log{\frac{z_0}{z}}}.
\end{equation}

And finally the metric is
\begin{eqnarray}
 ds^2 &=& {e^{-2\frac{y}{R}}}\left(\frac{y_0-y}{R}\right)^{\frac{1}{3b^2}} dx^\mu dx^\nu \eta_{\mu\nu} 
  - dy^2 \nonumber \\
 &=& \left(\frac{R}{z}\right)^2 \left(\left(\log{\frac{z_0}{z}}\right)^{\frac{1}{3b^2}}
  dx^\mu dx^\nu \eta_{\mu\nu} - dz^2 \right).
\end{eqnarray}

In order to write down the action of the dilaton couple into matter in Einstein frame, we compared the action in string and Einstein frame of a noncritical string theory in 5 dimensions~\cite{Csaki:2006ji}. In string frame the action has the form~\cite{Polchinski:1998rq, Batell:2008zm}
\begin{equation}
S = \frac{1}{2\kappa_0^2} \int d^5 x \sqrt{g_{str}}\left[ e^{-2 \Phi} (-1)\left(
\mathcal{R}_{str} +  4 \partial_{M} \Phi \partial^{M} \Phi+\tilde{V}(\Phi) \right) + e^{-\Phi} \mathcal{L}_{meson}\right]\, .
\label{nonstring}
\end{equation}
The action in Einstein frame is:
\begin{equation}
S = \frac{1}{2\kappa^2} \int d^5 x \sqrt{g} \left( -\mathcal{R} +\frac{4}{3} \partial_{M}
\Phi \partial^{M}\Phi -V(\Phi) + e^{\frac{7}{3}\Phi}\mathcal{L}_{meson}  \right) , \label{actionEinstein}
\end{equation}
where the metric of the string frame and the metric of the Einstein frame are related as $g_{MN}^{str}=e^{\frac{4}{3}\Phi}g_{MN}$. 
We can rescale the dilaton field to obtain a canonical kinetic term, by using $\Phi = \sqrt{3/8} \phi$.

%%%%%%%%%%%%%%%%%%%%%%%%%%%%%%%%%%%%%%%%%%%%
\section{Vector Meson}\label{sec:four}
%%%%%%%%%%%%%%%%%%%%%%%%%%%%%%%%%%%%%%%%%%%%
We have chosen the background dilaton profile such that the model produce asymptotic freedom. Here we will calculate the resulting vector meson spectrum on this background. Using the action in equation (\ref{actionEinstein}) and the rescaled dilaton field $\Phi = \sqrt{3/8} \phi$, the vector part of the action can be written as 
\be
S_V=\int d^5 x\, \sqrt{g}\,e^{\frac{7}{6}\sqrt{\frac{3}{2}}\phi} \left(\frac{-1}{4g_5^2}\right) g^{LM} g^{PN} F^{a}_{MN} F^{a}_{LP}\,, \label{actionvec}
\ee
with the following equation of motion for the bulk-to-boundary propagator of the transverse vector field
\be
\partial_z^2 V(q,z) -\left(1+\frac{-\frac{7}{6b}\sqrt{\frac{3}{2}}+\frac{1}{3b^2}}{\ln\frac{z_0}{z}}\right)\frac{\partial_z V(q,z)}{z} +\frac{q^2}{\left(\ln{\frac{z_0}{z}}\right)^{\frac{1}{3b^2}}}V(q,z)=0\,, \label{vecprop}
\ee
where $q$ denotes 4-momentum and $b$ as defined in (\ref{bphi}).

In order for  the action evaluated on the solution of the equation (\ref{vecprop}) to be finite when $b>(2/7)\sqrt{2/3}$, the bulk-to-boundary propagator or its derivative must vanish sufficiently fast at the IR boundary. 
%In fact, requiring $\partial_z V(q,z_0)=0$ automatically yields $V(q,z_0)=0$. 
Multiplication of both should vanish sufficiently fast such that the IR surface term of the action vanish. 

Evaluating the action on the solution, we obtain
\be
S_V= \int d^4 x \frac{1}{2g_5^2} V^0_\mu(x) P_T^{\mu\nu} V^0_\nu(x) \frac{R}{z} \left(\ln\frac{z_0}{z}\right)^{-\frac{7}{6b}\sqrt{\frac{3}{2}}+\frac{1}{3b^2}}V(q,z) \partial_z V(q,z)\bigg|_{z=\varepsilon\rightarrow 0}
\ee
where $P_T^{\mu\nu}=(\eta^{\mu\nu}-q^\mu q^\nu/q^2)$ and $V^0_\mu(x)$ is the UV boundary value of the vector field which corresponds to the source of vector current operator. In order to obtain 2-point function for the vector current, one has to take functional derivative with respect to the source $V^0_\mu$ twice on the 5D action. One obtains from the AdS/CFT correspondence
\be
i\int d^4 x \, e^{iqx} \left<0\bigg|\mathcal{T} J^{a\mu}(x)J^{b\nu}(0)\bigg|0\right>=-\frac{1}{g_5^2} P_T^{\mu\nu} \left(\ln\frac{z_0}{\varepsilon}\right)^{-\frac{7}{6b}\sqrt{\frac{3}{2}}+\frac{1}{3b^2}}\frac{\partial_z V(q,\varepsilon)}{\varepsilon}\,,\label{V2pointf}
\ee 
where we have set the AdS radius to unity ($R=1$). The bulk-to-boundary propagator can be normalized to $V(q,\varepsilon)=1$ at the UV boundary.

The vector meson mass spectrum can be obtained by solving the eigenvalue equation
\be
\partial_z^2 \psi_n(z) -\left(1+\frac{-\frac{7}{6b}\sqrt{\frac{3}{2}}+\frac{1}{3b^2}}{\ln\frac{z_0}{z}}\right)\frac{\partial_z \psi_n(z)}{z} +\frac{m_n^2}{\left(\ln{\frac{z_0}{z}}\right)^{\frac{1}{3b^2}}}\psi_n(z)=0\,. \label{vectoreigen}
\ee
where $m_n$ is the mass of the $n$-th Kaluza Klein modes. The wave function is normalized as follows
\be 
\int dz \,\frac{1}{z} \left(\ln \frac{z_0}{z}\right)^{-\frac{7}{6b}\sqrt{\frac{3}{2}}}\psi_n(z)\psi_m(z)=\delta_{mn}\,. \label{vecnorm}
\ee
The wave functions and their derivatives vanish at both the UV and the IR boundaries, except for the massless mode where the wave function approach a non-zero constant at IR boundary. The vanishing values of the wave functions at the boundaries is crucial for their normalizability.
%Except for the massless mode, 
%our numerical calculations indicate that 
%requiring the derivative of the wave function to vanish at IR boundary will make the value of the wave function itself to vanish there.
%This vanishing value of the wave function at this boundary is crucial for its normalizability. 

Setting $m_0=0$ in equation (\ref{vectoreigen}), one obtains the following solution
\be
\psi_0=N_0 \Gamma\left(1-\frac{1}{3b^2}+\frac{7}{6b}\sqrt{\frac{3}{2}}, 2 \ln\frac{z_0}{z}\right) \,,
\ee
which is the upper incomplete gamma function. This function is vanishing at UV boundary and approach $N_0\Gamma(1-1/(3b^2)+7/(6b)\sqrt{3/2})$ at IR boundary.  The slope vanishes at the IR boundary for $b>(2/7)\sqrt{2/3}$. The weighting function of the normalization integral is infinite at both UV and IR boundary. However, the vanishing value of $\psi_0$ near UV overcome this to make the integrand in (\ref{vecnorm}) goes to zero at the UV boundary. This is not the case at the IR boundary as the wave function goes to a non-zero constant when $b>(2/7)\sqrt{2/3}$. Near the IR boundary, one can pull out $\psi_0^2$ out of the integral. This leave us with the following integral
\be
\int_\eta^{z_0} dz \frac{1}{z} \left(\ln\frac{z_0}{z}\right)^{-\frac{7}{6b}\sqrt{\frac{3}{2}}}
\ee
which is infinite for $0<b<(7/6)\sqrt{3/2}$, with $\eta$ close to $z_0$. In the next section we will show that with $b>0$ the quark mass runs properly, that is, in the high energy limit as the energy scale increases, the quark mass decreases as in QCD.

%For definiteness we set $b=\sqrt{2/3}$ and for the massless mode, we obtain
%\be
%\psi_0=N_0 \Gamma\left(\frac{9}{4}, 2 \ln\frac{z_0}{z}\right) \label{vecnorm}\,,
%\ee
%which is the upper incomplete gamma function. This function is vanishing at UV boundary and approach $\Gamma(9/4)$ with vanishing slope at the IR boundary. The weighting function of the normalization integral is infinite at both UV and IR boundary. However, the vanishing value of $\psi_0$ near UV overcome this to make the integrand in (\ref{vecnorm}) goes to zero at the UV boundary. This is not the case at the IR boundary as the wave function goes to a non-zero constant. Near the IR boundary, one can pull out $\psi_0^2$ out of the integral. Evidently,  $\psi_0$ normalization integral is infinite as the integral
%\be
%\int_\eta^{z_0} dz \frac{1}{z} \left(\ln\frac{z_0}{z}\right)^{-\frac{7}{4}}
%\ee
%is infinite for some $\eta$ close to $z_0$.  Therefore, we discard the massless mode. 

%%%%%%%%%%%%%%%%%%%%%%%%%%%%%%%%%%%%%
\begin{figure}
\centering
\includegraphics[width = 2.5in]{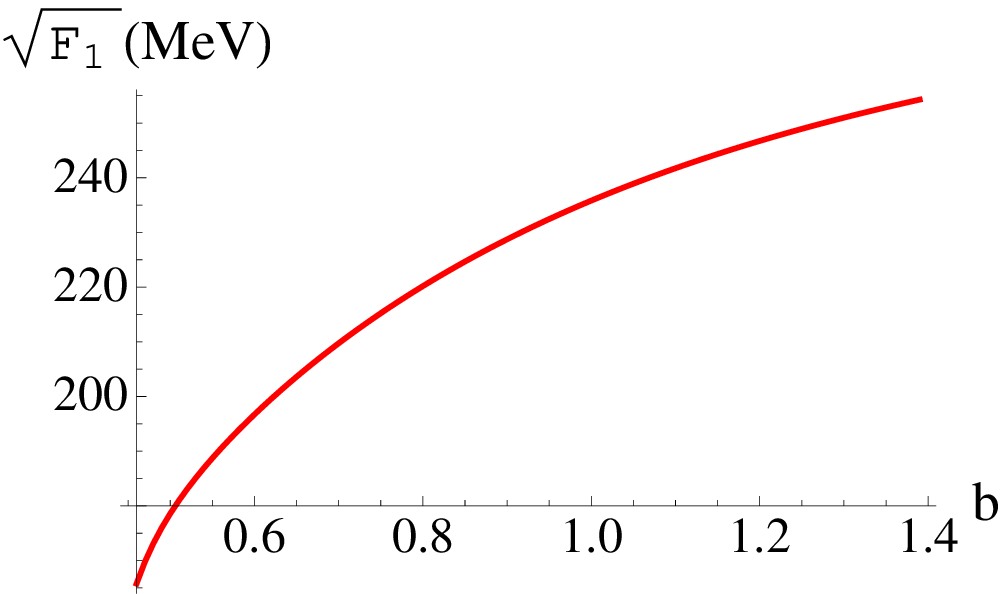}\includegraphics[width = 2.5in]{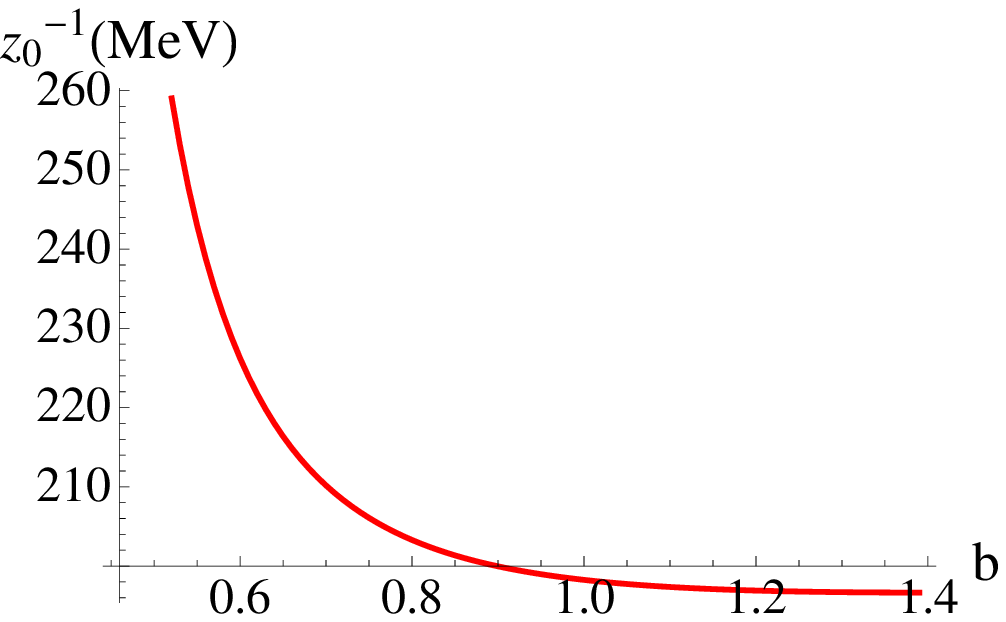}
\caption{Left: Plot of vector meson's decay constant $F_1^{1/2}$ as a function of $b$. Right: Plot of $z_0^{-1}$ as a function of $b$. In both plot we fix the lightest vector meson's mass to $775.5$ MeV}
\label{fig:fvz0r}
\end{figure}
%%%%%%%%%%%%%%%%%%%%%%%%%%%%%%%%%%%%%

In the case of $b=\sqrt{2/3}$, the first four mass eigenvalues of equation  (\ref{vectoreigen}) ( in unit of $1/z_0$) are $3.83$, $6.52$, $9.14$, $11.74$. We set the value of $z_0$ to fit the mass of the first mode to the mass of rho meson. For $m_\rho=775.5$ MeV, we obtain $1/z_0=202.6$ MeV. The resulting mass of the first few excited states are $1321$ MeV, $1852$ MeV, $2379$ MeV. This mass spectrum grows like $m_n^2\sim n^2$, similar to the hard-wall model. As comparisons, in the original hard-wall model, the corresponding mass of the excited states are $1780$ MeV, $2789$ MeV, $3802$ MeV. For the soft-wall model with quadratic dilaton profile, the corresponding mass are $1097$ MeV, $1343$ MeV, $1551$ MeV. 

The bulk-to-boundary propagator can be written as a sum over normalizable modes,
\be
V(q,z)=-\sum\frac{\left(\frac{1}{z}\left(\ln\frac{z_0}{z}\right)^{-\frac{7}{6b}\sqrt{\frac{3}{2}}+\frac{1}{3b^2}}\partial_z\psi_n\right)_{\varepsilon} \psi_n(z)}{q^2-m_n^2}\,.
\ee
This expansion can be substituted into (\ref{V2pointf}) from which one can identify the decay constant of the vector meson
\be
F_n=\frac{1}{g_5\varepsilon}\left(\ln\frac{z_0}{\varepsilon}\right)^{-\frac{7}{6b}\sqrt{\frac{3}{2}}+\frac{1}{3b^2}} \partial_z\psi_n(\varepsilon)\,,
\ee
where $g_5=2\pi$~\cite{Erlich:2005qh}. We obtain $F_1=(222\, {\rm MeV})^2$. As comparisons, the hard-wall model and the soft-wall model with quadratic dilaton profile yield $F_1=(329\, {\rm MeV})^2$ and $F_1=(260\, {\rm MeV})^2$, respectively.

We can vary the value of $b$ while fixing the mass of the lightest vector meson to $775.5$ MeV. The results for the decay constant and the value of $z_0^{-1}$ are shown in Figure \ref{fig:fvz0r}. 
%%%%%%%%%%%%%%%%%%%%%%%%%%%%%%%%%%%%%%%%%%%%

%%%%%%%%%%%%%%%%%%%%%%%%%%%%%%%%%%%%%
\begin{figure}
\centering
\includegraphics[width=15 cm]{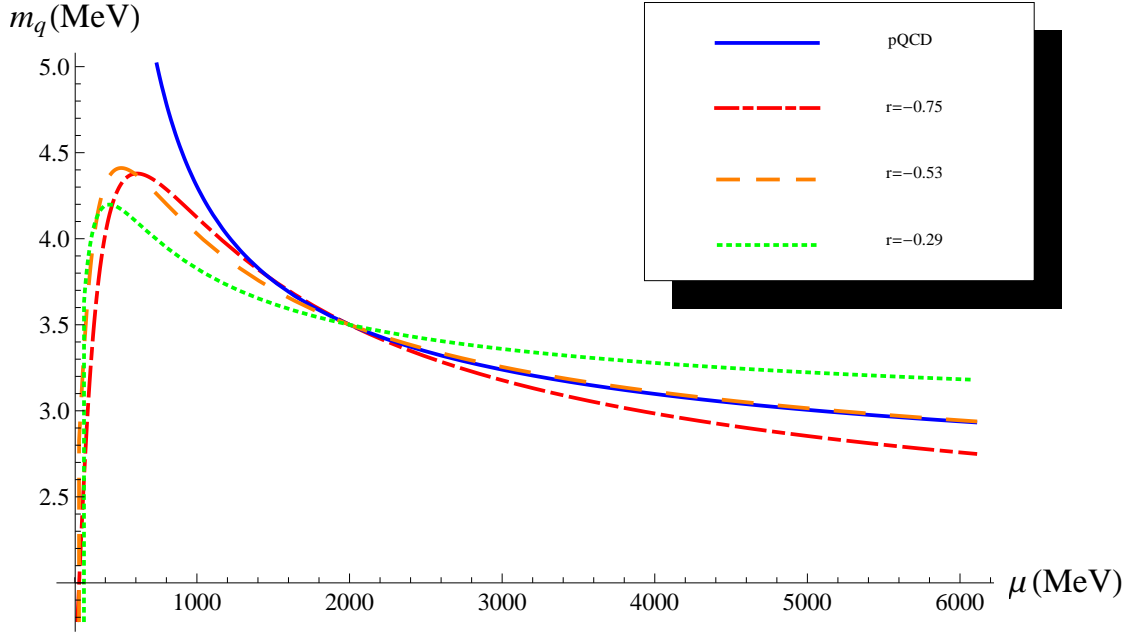}
\caption{(Color online). Plot of running of the quark mass as a function of energy scale for different $r$. The blue ({\color{blue}---}) is the running of quark mass from QCD. The green ({\color{green}.......}) is for $r=-0.29\,(b=0.52)$, the orange ({\color{BurntOrange}--- --- ---}) is for $r=-0.53\, (b=0.60)$, and the red ({\color{red}--- -- --- -- ---}) is for $r=-0.75\, (b=\sqrt{2/3})$.}
\label{mqrun}
\end{figure}
%%%%%%%%%%%%%%%%%%%%%%%%%%%%%%%%%%%%%

%%%%%%%%%%%%%%%%%%%%%%%%%%%%%%%%%%%%%
\begin{figure}
\centering
\includegraphics[width=15 cm]{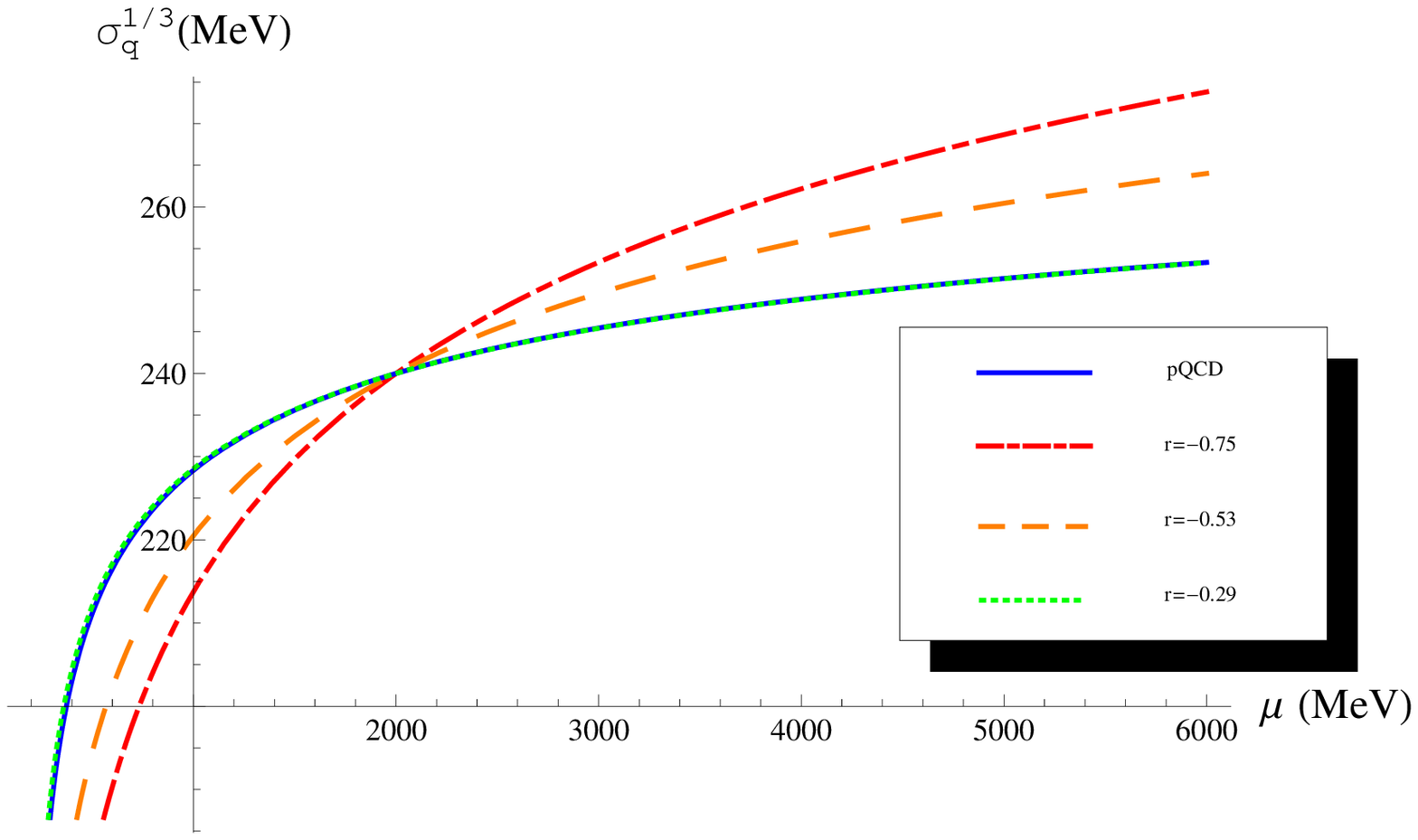}
\caption{(Color online). Plot of running of the quark condensate as a function of energy scale for different $r$. The blue ({\color{blue}---}) is the running of quark mass from QCD. The green ({\color{green}.......}) is for $r=-0.29 \,(b=0.52)$, the orange ({\color{BurntOrange}--- --- ---}) is for $r=-0.53\, (b=0.60)$ and , the red ({\color{red}--- -- --- -- ---}) is for $r=-0.75\, (b=\sqrt{2/3})$. }
\label{sigrun}
\end{figure}
%%%%%%%%%%%%%%%%%%%%%%%%%%%%%%%%%%%%%

%%%%%%%%%%%%%%%%%%%%%%%%%%%%%%%%%%%%%%%%%%%%
%
\section{Running quark mass and QCD condensate}\label{sec:three}
%
%%%%%%%%%%%%%%%%%%%%%%%%%%%%%%%%%%%%%%%%%%%%

The action for dilaton coupled to matter in Einstein frame is:
\begin{equation}
 {\cal S} = \int d^5x \sqrt{g} e^{\frac{7}{6}\sqrt{\frac{3}{2}}\phi} 
  {\rm Tr} \lbrace |DX| + 3|X|^2 - \frac{1}{4g_5^2} (F_L^2 + F_R^2) \rbrace,
\end{equation}
with $D^M X = \partial^M X - iA_L^M X + i XA_R^M$ and $F_{L,R}^{MN} = 
\partial^M A_{L,R}^N - \partial^N A_{L,R}^M - i[A_{L,R}^M,A_{L,R}^N]$.

The equation of motion for the background field $X_0$ in the $u=(y_0-y)/R$ coordinate:
\be
\partial_u\left( e^{4u} (u)^{r}\partial_u X_0\right)+3 e^{4u} (u)^{r} X_0 =0,
\label{scalareom}
\ee
where $r = \frac{2}{3b^2} -\frac{7}{6b} \sqrt{\frac{3}{2}}$.   
The solutions of the equation of motion (\ref{scalareom}) from~\cite{Olver:2010:N,AbramowitzS} in terms of Kummer functions are:
\be
X_0(u) =C_1 e^{-3 u} u^{1-r} M\left[1+\frac{r}{2},2-r,2u\right] + C_2 e^{-3 u} u^{1-r} U\left[1+\frac{r}{2},2-r,2u\right] ,
\label{eqnxu}
%X(u) = C_1 e^{-3 u} u^{1-r} U\left[1+\frac{r}{2},2-r,2u\right] +C_2 e ^{-u}u^{1-r} U\left[1-\frac{3r}{2},2-r,-2u\right] 
\ee
with $C_1$ and $C_2$ are integration constants.  Contrary to the soft-wall model, we can keep both terms since the action evaluated on the above solution remains finite.

In order to understand how the scalar field behaves near the UV boundary, we write the solution in the following form, using coordinate transformation $u = -\ln (z/z_0)$ and Kummer's transformation~\cite{Olver:2010:N,AbramowitzS}
\be
X_0(z) =\tilde{C}_1 (\ln\frac{z_0}{z})^{1-r} U\left[1-\frac{3r}{2},2-r,2\ln\frac{z}{z_0}\right] \left(\frac{z}{z_0}\right)
+\tilde{C}_2 (\ln\frac{z_0}{z})^{1-r} U\left[1+\frac{r}{2},2-r,-2\ln\frac{z}{z_0}\right] \left(\frac{z}{z_0}\right)^3.
\label{eqnxz}
\ee
The constants are related as follows
\ba
\tilde{C}_1&=&-C_1\frac{\Gamma(2-r)}{\Gamma(1+r/2)}\exp(-i3\pi r/2)\,,\\
\tilde{C}_2&=&C_2-C_1\frac{\Gamma(2-r)}{\Gamma(1-3r/2)}\exp(-i\pi r/2)\,.
\ea
The Kummer's functions in equation (\ref{eqnxz}) can be written as an infinite series
\ba
X_0(z)&=&\tilde{C}_1\,(-2)^{(3r/2-1)} \left(\frac{z}{z_0}\right)\left(\ln \frac{z_0}{z}\right)^{\frac{r}{2}}\sum_{n=0}^\infty\frac{\left(1-\frac{3r}{2}\right)_n\left(-\frac{r}{2}\right)_n}{2^n n!}\left(\frac{1}{\ln\frac{z_0}{z}}\right)^n \nonumber\\
&&+\tilde{C}_2\, (2)^{-(1+r/2)}\left(\frac{z}{z_0}\right)^3\left(\ln \frac{z_0}{z}\right)^{-\frac{3r}{2}}\sum_{n=0}^\infty\frac{\left(1+\frac{r}{2}\right)_n\left(\frac{3r}{2}\right)_n}{2^n n!}\left(\frac{-1}{\ln\frac{z_0}{z}}\right)^n\,,
%&=&\left(\frac{z}{z_0}\right)\left(\ln \frac{z_0}{z}\right)^{\frac{r}{2}}\left[1- \frac{r}{4}\left(1-\frac{3r}{2}\right)\left(\frac{1}{\ln\frac{z_0}{z}}\right)+\ldots\right]\nonumber\\
%&&+\left(\frac{z}{z_0}\right)^3\left(\ln \frac{z_0}{z}\right)^{-\frac{3r}{2}}\left[1- \frac{3r}{4}\left(1+\frac{r}{2}\right)\left(\frac{1}{\ln\frac{z_0}{z}}\right)+\ldots\right]
\ea
where $(a)_n=(a)(a+1)\ldots(a+n-1)$, and $(a)_0=1$.  In the limit $z\to 0$ both term in above equation vanish, however, the second term vanish faster than the first term because of the $z^3$ factor. Hence, in this limit we can drop the second term and identify the remaining term as quark mass multiplying $\zeta z/2$ as in (\ref{mqsigEq}). 

%Based on~\cite{Erlich:2005qh}, the value of field $X$ corresponding to the quark mass $m_q$ and QCD condensate $\sigma$:
%\be
%X_0(z) = \dfrac{1}{2} m_q\frac{\sqrt{N_c}}{2\pi} z +\dfrac{1}{2} \sigma \frac{2\pi}{\sqrt{N_c}} z^3.
%\ee

We will determine the constants of integration using the value of $m_q$ and $\sigma$ at $1/z =\unit[2]{GeV}$ from~\cite{Nakamura:2010zzi} as input parameter. 
%In order to extract the quark mass and QCD condensate we transform the solutions by using Kummer's transformation~\cite{Olver:2010:N,AbramowitzS} and coordinate transformation $u = -\ln (z/z_0)$:
%\be
%X_0(z) =\tilde{C}_1 (-\ln\frac{z}{z_0})^{1-r} U\left[1-\frac{3r}{2},2-r,2\ln\frac{z}{z_0}\right] z
%+\tilde{C}_2 (-\ln\frac{z}{z_0})^{1-r} U\left[1+\frac{r}{2},2-r,-2\ln\frac{z}{z_0}\right] z^3.
%\label{eqnxz}
%\ee
Because the imaginary parts of equation (\ref{eqnxz}) cancel out as a consequence of  equation (\ref{eqnxu}) which is real, we take the real part of $X_0(z)$ and associate them to the quark mass $m_q$ and QCD condensate $\sigma$:
\begin{align}
m_q (z) &= \mathrm{Re}\left[\tilde{C}_1 (\ln\frac{z_0}{z})^{1-r} U\left[1-\frac{3r}{2},2-r,2\ln\frac{z}{z_0}\right] \right] \frac{4\pi}{\sqrt{N_c}} \label{mq} ,\\
\sigma (z) &=  \mathrm{Re}\left[\tilde{C}_2 (\ln\frac{z_0}{z})^{1-r} U\left[1+\frac{r}{2},2-r,-2\ln\frac{z}{z_0}\right] \right]\frac{\sqrt{N_c}}{\pi}\label{sig}. 
\end{align}

By associating $z_0$ with $1/\Lambda_{QCD}$, equation (\ref{mq}) and (\ref{sig}) show how the  running of quark mass and QCD condensate as a function of energy scale $1/z = \mu$. Figure (\ref{mqrun}) shows the running of the quark mass from equation (\ref{mq}) compared with the running of quark mass from perturbative QCD up to two loops calculation~\cite{Koide:1994au,Fusaoka:1998vc}, 
\begin{equation}
m_q(\mu)=\widetilde{m}_q\left(\frac{1}{2}L\right)^{-2\gamma_0/\beta_0}
\left[1-\frac{2\beta_1\gamma_0}{\beta_0^3}
\frac{\ln L+1}{L}+\frac{8\gamma_1}{\beta_0^{2}L}+O(L^{-2}\ln^2 L)\right],\label{pQCD}
\end{equation}
where $\gamma_0=2$, $\gamma_1=\frac{101}{12}-\frac{5}{18}n_q$,  $\beta_0=11-\frac{2}{3}n_q$, $ \beta_1=51-\frac{19}{3}n_q$, and $L=\ln(\mu^2/\Lambda^2)$.
Here, $\widetilde{m}_q$ is the renormalization group invariant mass. 

We vary the value of $r$ while fixing the mass of the lightest vector meson $m_\rho=\unit[757]{MeV}$. The best fit as shown in figure (\ref{mqrun}) is for $r = -0.53$ which corresponds to $1/z_0=\unit[226]{MeV}$ and $\sqrt{F_\rho}=197$ MeV.

In figure (\ref{sigrun}), we fit equation (\ref{sig}) to the one loop QCD calculation for the running of quark condensate in the chiral limit~\cite{Williams:2007ef,Williams:2007ey}:
\begin{equation}
	\left<\overline{q}q\right>(\mu) = \left( \frac{1}{2}\log\frac{\mu^2}{\Lambda^2} \right)^{\gamma_m}\left<\overline{q}q\right>\,,
\end{equation}
with $\gamma_m=12/(33-2n_q)$. The best fit is for $r=-0.29$ which corresponds to $1/z_0 = \unit[259]{MeV}$ and $\sqrt{F_\rho}=183$ MeV.
%%%%%%%%%%%%%%%%%%%%%%%%%%%%%%%%%%%%%%%%%%%%
\section{Conclusion}\label{sec:concl}
%%%%%%%%%%%%%%%%%%%%%%%%%%%%%%%%%%%%%%%%%%%%
In this paper we have calculated the running of the quark mass and the quark condensate using a dynamical soft-wall model. In order to remove the massless vector meson and to make the quark mass runs properly we have determined the allowed value of parameter $b$ to be $0<b<(7/6)\sqrt{3/2}$. We also find that, contrary to the soft-wall model with quadratic dilaton profile, the model allows for a non-vanishing quark condensate in the chiral limit. We obtain the vector meson mass spectrum similar to that of the hard-wall model. The best fit of this model to the running of the quark mass yield $r=-0.53$, which corresponds to $b=0.60$ and the best fit to the running of the quark condensate yield $r=-0.29$, which corresponds to $b=0.52$.

%%%%%%%%%%%%%%%%%%%%%%%%%%%%%%%

%\begin{acknowledgments}

%\end{acknowledgments}

%%%%%%%%%%%%%%%%%%%%%%%%%%%%%%%

%\bibliographystyle{amsplain}
\bibliographystyle{JHEP}

\bibliography{adscft}

\end{document}